# Overtaking-enabled Eco-approach Control at Signalized Intersections for Connected and Automated Vehicles

Haoxuan Dong, Weichao Zhuang, *Member, IEEE*, Guoyuan Wu, *Senior Member, IEEE*, Zhaojian Li, *Senior Member, IEEE*, Guodong Yin, *Senior Member, IEEE*, Ziyou Song, *Member, IEEE*

***Abstract*—Preceding vehicles typically dominate the movement of following vehicles in traffic systems, thereby significantly influencing the efficacy of eco-driving control that concentrates on vehicle speed optimization. To potentially mitigate the negative effect of preceding vehicles on eco-driving control at the signalized intersection, this paper proposes an overtaking-enabled eco-approach control (OEAC) strategy. It combines driving lane planning and speed optimization for connected and automated vehicles to relax the first-in-first-out queuing policy at the signalized intersection, minimizing the target vehicle's energy consumption and travel delay. The OEAC adopts a receding horizon two-stage control framework to derive optimal driving trajectories for adapting to dynamic traffic conditions. In the first stage, the driving lane optimization problem is formulated as a Markov decision process and solved using dynamic programming, which takes into account the uncertain disturbance from preceding vehicles. In the second stage, the vehicle's speed trajectory with the minimal driving cost is optimized rapidly using Pontryagin's minimum principle to obtain the closed-form analytical optimal solution. Extensive simulations are conducted to evaluate the effectiveness of the OEAC. The results show that the OEAC is excellent in driving cost reduction over constant speed and regular eco-approach and departure strategies in various traffic scenarios, with an average improvement of 20.91% and 5.62%, respectively.**

## I. Introduction

The growth of urbanization and the increasing vehicle ownership have led to an increase in the number of vehicles in urban traffic [1, 2], raising concerns about greenhouse gas emissions and traffic congestion [3]. Fortunately, by connecting vehicles and road infrastructure, the emerging connected and automated vehicle (CAV) technologies provide possibilities to further improve vehicle energy efficiency and traffic throughput [4]. In this context, a variety of eco-friendly methods have been proposed from road or vehicle perspectives to improve vehicle and traffic efficiency. One such method is to optimize the traffic light signal phase and timing (SPaT) for prioritization of a single vehicle or a platoon approaching the intersection [5]. Another approach involves economically controlling the vehicle speed drive to pass through the signalized intersection, known as eco-approach and departure (EAD) control in some literature [6].

The EAD aims to facilitate smooth vehicle passage through signalized intersections, by avoiding stop-and-go behavior, and utilizing look-ahead information on traffic lights and preceding vehicles [6, 7]. A simple EAD is green light optimized speed advisory system, which uses deterministic traffic light SPaT, speed limits, and vehicle kinematics to calculate the speed range that ensures the vehicle can pass through an intersection without stopping [8]. Considering the uncertain SPaT of actuated traffic lights, Mahler *et al.* [9] introduced a prediction model to determine the probability of green signals at each time step, then the vehicle energy-optimal speed trajectory was derived that maximizes the likelihood of passing through an intersection. In real-world traffic, the vehicle speed may be impacted by the preceding vehicle or the queue waiting at the intersection. Therefore, Hao *et al.* [7], Bai *et al.* [10], and Ye *et al.* [11] designed the EAD strategies considering preceding vehicle movement information obtained using on-board radar, vehicle-to-vehicle communication, or prediction model. Dong *et al.* [12] and Sun *et al.* [13] proposed enhanced EAD to allow the vehicle to drive through the intersection without a stop by considering the discharging time of the vehicle queue, where the vehicle kinematics and traffic shockwave theory were utilized to estimate the vehicle queue movement. Other EAD strategies were proposed by Wang *et al.* [14], Han *et al.* [15], and Dong *et al.* [16]. These strategies are based on designing multiple intersection-based EAD controller that incorporates the spatial-temporal correlation among signalized corridors. Ma *et al.* [17], Li *et al.* [18], and Jiang *et al.* [19] proposed vehicle platoon-based EAD approaches that improve the energy efficiency of multiple vehicles passing through the signalized intersection.

Most EAD research has focused on optimizing vehicle speed to control its longitudinal movement. However, in some scenarios, the host vehicle (HV) may encounter slow-

Manuscript received April 11, 2023. This work was supported by the Energy Storage Lab at National University of Singapore. (*Corresponding author: Ziyou Song.*)

H. Dong and Z. Song are with the Department of Mechanical Engineering, National University of Singapore, Singapore 117575, Singapore (e-mail: donghaox@foxmail.com; ziyou@nus.edu.sg).

W. Zhuang and G. Yin are with the School of Mechanical Engineering, Southeast University, Nanjing 211189, China (e-mail: wezhuang@seu.edu.cn; ygd@seu.edu.cn).

G. Wu is with the Department of Electrical and Computer Engineering, University of California, Riverside, Riverside, CA 92507, USA (e-mail: guoyuan.wu@ucr.edu).

Z. Li is with the Department of Mechanical Engineering, Michigan State University, East Lansing, MI 48824, USA (e-mail: lizhaoj1@egr.msu.edu).

moving vehicles ahead. If the HV does not overtake the preceding vehicle, the HV speed optimization is restricted to the car-following mode, resulting in degraded efficacy of EAD control. Earlier vehicle overtaking research concentrated on lane-changing decision and trajectory optimization, Zhou *et al.* [20] and Li *et al.* [21] formulated optimal lane-changing strategies that consider HVs and other vehicles in their vicinity to improve driving comfort, safety, and lane-changing efficiency in complex driving environments. These studies, however, omit the vehicle energy-saving objective, which could lead to a reduction in vehicle energy efficiency when frequent lane changes are present. To address this issue, Gu *et al.* [22] have developed a joint control approach to simultaneously regulate vehicle longitudinal and lateral movement, allowing CAVs to travel in an energy-saving manner by adjusting the space headway and driving speed. Chen *et al.* [23] have derived time and energy-optimal lane-changing control policies that cooperate with neighboring CAVs.

The studies mentioned above focus on overtaking-enabled eco-cruising control on highways; however, the offered solutions may not apply to urban traffic due to the presence of traffic lights. Lane-changing in urban traffic is not only for achieving the goals of improving vehicle energy efficiency and reducing travel delay by considering the impact of preceding vehicles as in a highway scenario, but also to ensure vehicles pass through signalized intersections without stop-and-go operation during the green window. Currently, there are few investigations of overtaking-enabled EAD. Yang *et al.* [24] designed a less-disturbed EAD strategy that integrates offline energy-efficient speed planning and online speed tracking with overtaking ability. Nevertheless, the less-disturbed EAD does not achieve an energy-saving control globally since lane changing is not considered in vehicle speed planning. Guo *et al.* [25] developed a hybrid deep Q-learning and policy gradient-based EAD under a full penetration rate of CAVs environment, where both speed optimization and lane-changing order were considered. However, because a fully connected environment can only be realized in the distant future [26, 27] and the proposed approach does not take vehicle lane-changing trajectory into consideration, this approach is not practical in real-world traffic. Moreover, Hu *et al.* [28] proposed an enhanced EAD controller to cut through the traffic and catch green signals by overtaking slow-moving vehicles for real-time application in a partially connected environment. However, this study formulated a nonlinear optimal control problem based on a simplified vehicle model that ignored the vehicle powertrain model. The resulting solution could be different from reality.

This study proposes an overtaking-enabled eco-approach control (OEAC) strategy for CAVs, which is capable of relaxing the first-in-first-out (FIFO) queuing policy at a signalized intersection, while minimizing vehicle energy consumption and travel delay in a computationally efficient manner. The major contributions are threefold:

1) An overtaking-enabled EAD optimal control problem is formulated to cooperate with driving lane planning and speed optimization while considering the constraints posed by traffic lights and preceding vehicles. The unified monetary counterpart objectives of energy consumption and travel time are used to balance vehicle energy and traffic efficiency.

2) A real time optimal driving trajectory is derived through a receding horizon two-stage control framework. In the first stage, the driving lane decision is formulated as a Markov decision process (MDP), and dynamic programming is adopted to solve this problem while considering the uncertain disturbance of preceding vehicles. In the second stage, the speed trajectory is optimized using Pontryagin's minimum principle (PMP) algorithm with minimal driving costs, leading to rapid optimization.

3) The simulation of stochastic traffic scenarios and typical case studies are investigated to verify the effectiveness of the proposed OEAC strategy. Furthermore, the sensitivity of OEAC's effectiveness to traffic flow conditions and traffic light initial states is inspected.

The remainder of this paper is organized as follows. Section II formulates the intersection crossing scenario and vehicle model. In Section III, the EAD optimal control problem and the OEAC framework are defined. Section IV provides the methodology of the OEAC in detail. The performance of the OEAC is comprehensively evaluated in Section V. Finally, Section VI concludes this paper.

## II. Intersection Crossing Scenario and Vehicle Model

### *A. Intersection Crossing Scenario*

We consider a general three-lane urban route with a fixed-timing traffic light, as shown in Fig. 1.

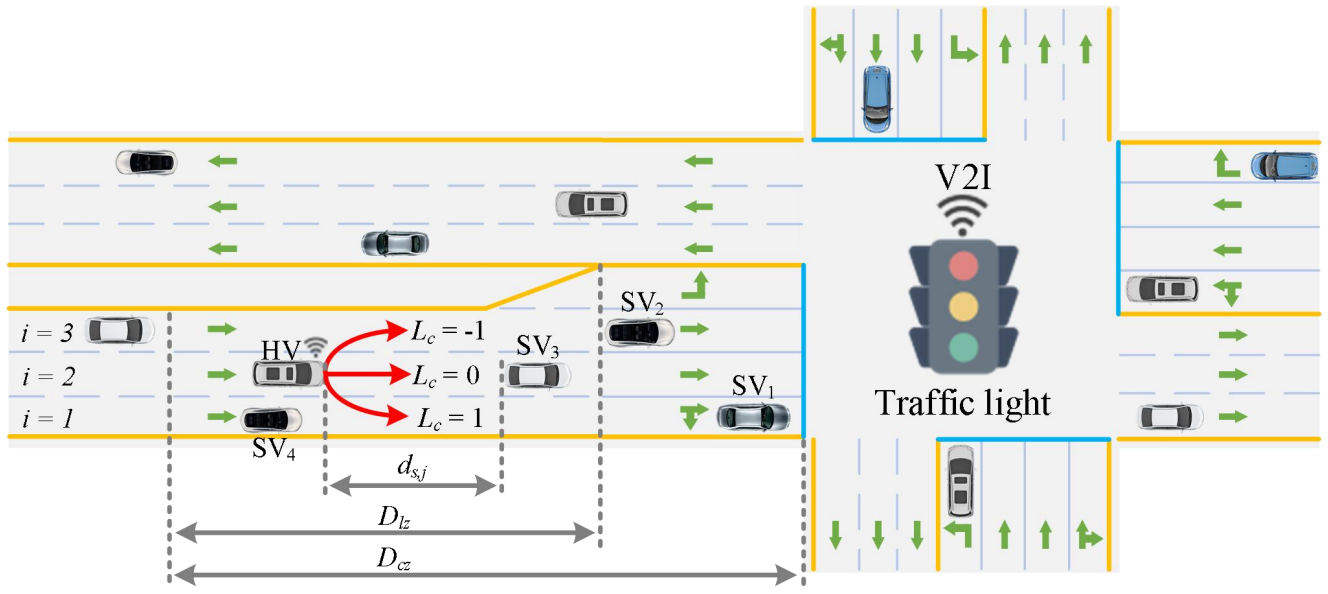

**Fig. 1.** A general signalized intersection route.

We define the route as a set $O_r = \{S_{tl}, D_{cz}, D_{lz}, N_l, D_w, V_{max}, V_{min}\}$. $S_{tl}$ is the location of the stop line at the intersection, $D_{cz}$ is the communication zone, $D_{lz}$ is the lane-changing coordinator zone, which is related to $D_{cz}$ and the length of no lane-changing zone near the intersection stop line. $N_l$ is the total number of lanes in the same driving direction as the HV, where the lane numbers from the outside to the inside of the road are $i = 1, 2, \cdots, N_l$. $D_w$ is the width of the lane, and $V_{max}$ and $V_{min}$ are maximum and minimum speed limits, respectively.

The information on traffic lights is defined by a set $O_t = \{T_{ls}, I_{in}, T_{gr}, T_{re}\}$, where $T_{ls}$ is the initial transition time of the traffic light indication when the HV is approaching the communication zone, $I_{in}$ is the initial indication of the traffic light with $I_{in} = 1$ and $I_{in} = 0$ denoting the green and red signals, respectively, and $T_{gr}$ and $T_{re}$ are the time interval of green and red signals, respectively. Note that the yellow signal is lumped into the red for driving safety concerns.

The set of surrounding vehicles (SV) is given by $\mathcal{O}_s = \{N_{sv}, s_{s,j}, v_{s,j}, L_{s,j}\}$. Specifically, $N_{sv}$ is the number of SVs, where we specify vehicle serial numbers as $j = 1, 2, \cdots, N_{sv}$, in the order of proximity to the intersection and being from the innermost lane to outermost lane. $s_{s,j}$, $v_{s,j}$, and $L_{s,j}$ are the position, speed, and driving lane of $j$th SV, respectively. The SVs are classified into three types by their relative position with respect to the HV in the longitudinal direction: the vehicles in front of the HV are the preceding vehicles, the vehicles driving side by side with the HV are the side vehicles, and the vehicles behind the HV are the rear vehicles.

It is also assumed that the HV is equipped with a vehicle-to-infrastructure communication device (4G cellular network) and can access instantaneous traffic information. The HV can modify its speed in the communication zone and change the driving lane in the lane change coordinator zone. Note that the lane-changing rule is constructed to only permit the HV to drive in the current lane or change to the adjacent right or left lanes, as shown in Fig. 1. Therefore, the maximum permissible lane shift is only one, and the lane-changing index $L_c$ has only three values, i.e., $L_c = [-1, 0, 1]$, corresponding to changing lanes to the left, driving in the current lane, and changing lanes to the right, respectively.

## B. Vehicle Model

When changing lanes, we assume that the steering action is mild and the speed is moderate, so the vehicle complies with the dynamics and movement geometry constraints. Then, the bicycle model with an Ackerman steering design is used to explain vehicle lateral and longitudinal movements [28], as expressed in (1),

$$\begin{bmatrix} \dot{s_x} \\ \dot{s_y} \\ \dot{\gamma} \end{bmatrix} = \begin{bmatrix} \cos\gamma \\ \sin\gamma \\ \tan\beta_w / L \end{bmatrix} v \tag{1}$$

where $v$ is the speed of HV, $\gamma$ is the yaw angle, $\beta_w$ is the front-wheel steering angle, $L$ is the HV body length, and $s_x$ and $s_y$ are the HV's longitudinal and lateral positions, respectively.

We focus on the EAD control of an electric vehicle that is powered by a centralized electric motor. Ignoring the electric wire loss, the battery power is calculated by (2),

$$P_b = P_m \eta_b^{-\text{sign}(P_m)} + P_a \eta_b^{-1} \tag{2}$$

where $\eta_b$ is the battery efficiency, $P_a$ is the vehicle accessory power, and $P_m$ is the motor power. The approximated closed-form expression of motor power [29] is used to reduce the computational burden, as expressed in (3),

$$P_m = \varpi T_m v + \kappa T_m^2 \tag{3}$$

where $\varpi$ and $\kappa$ are the empirical coefficients, i.e., $\varpi = i_g i_0 / r_w$ and $\kappa = R_m / \mu^2$. $R_m$ is the motor armature resistance, $\mu$ is the motor speed constant, and $T_m$ is the motor torque, as defined in (4) [30],

$$T_m = \frac{(mgf\cos\theta + mg\sin\theta + 0.5C_D A\rho v^2 + m\delta\dot{v})r_w}{i_g i_0 \eta_t^{\text{sign}(T_m)}} \tag{4}$$

where $m$ is the vehicle mass, $g$ is the gravity acceleration, $f$ is the rolling resistance coefficient, $C_D$ is the aerodynamic drag coefficient, $A$ is the frontal area, $\rho$ is the air density, $\delta$ is the vehicle rotational inertia coefficient, $r_w$ is the radius of the vehicle tire, $i_g$ is the transmission ratio, $i_0$ is the final drive ratio, $\eta_t$ is the driveline efficiency, $\theta$ is the road slope, and sign$(\cdot)$ is the signum function.

# III. Problem Formulation and Control Framework

## A. EAD Problem Analysis

EAD control is recognized as a promising approach to reducing vehicle energy consumption and travel delay in urban traffic at signalized intersections [4]. Regular EAD (READ) uses traffic light and preceding vehicle information, as shown in the blue solid line in Fig. 2, to calculate the vehicle speed to pass through the signalized intersection without stop-and-go operation. While approaching the preceding vehicle, the HV must adhere to the FIFO queuing policy, which may increase the travel time when encountering a slow-moving preceding vehicle. Additionally, due to the blocking effect of the preceding vehicle, the speed of READ may not be optimal.

In this context, to realize a more efficient vehicle control system, this paper proposes an overtaking-enabled EAD strategy, referred to as OEAC, for safely and efficiently passing through a signalized intersection by combining driving lane planning and speed optimization. As shown in the green solid line in Fig. 2, the HV with OEAC can overtake the slower vehicle, therefore relaxing the FIFO queuing policy to lower vehicle energy consumption and travel delay effectively.

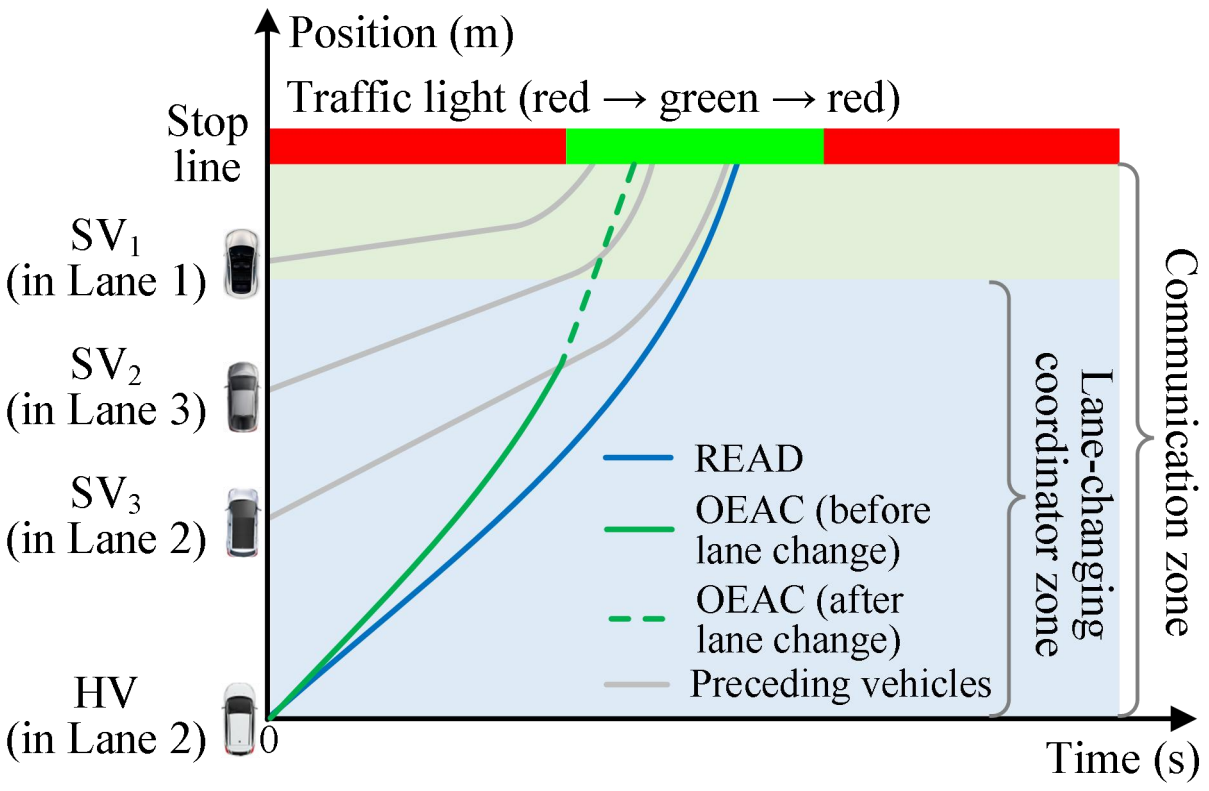

**Fig. 2.** Schematic diagram of a vehicle crossing the signalized intersection with READ and OEAC strategies. Here, the HV initially drives in Lane 2 before changing to Lane 1.

## B. Optimization Control Problem

The objective of the proposed OEAC is to maneuver the vehicle efficiently passing through the signalized intersection with a minimal driving cost. The controller must determine the optimal driving lane and speed for the HV, considering relevant vehicle and traffic constraints, to reach the destination with the shortest amount of time and reduced energy consumption.

We define the state variable $x$ as including vehicle speed and longitudinal position, i.e., $x = [v, s_x]^T$. To improve vehicle energy efficiency, we assume no mechanical brake force is actuated [31], then the control input $u$ includes motor torque and lane-changing index, which is defined as $u = [T_m, L_c]^T$ .Suggestion To mitigate the potential conflict between two goals: the lowest energy consumption and shortest travel time [16], a monetary counterpart normalized objective function of vehicle energy and travel time is established to calculate the driving cost. Consequently, the overtaking-enabled EAD optimal control problem is formulated in (5) and (6), with the goal to minimize driving cost (see (5)), under muti-constraints, i.e., speed limits (see (6a)), motor torque bounds (see (6b)), lane change integer constraint (see (6c)), safety inter-vehicle distance (see (6d)), and traffic light control policy (see (6e)),

$$\min_{u \in U} \mathcal{J}(x,u) = \int_{T_s}^{T_f} (\zeta_e P_b + \zeta_t) dt \tag{5}$$

subject to:
$$V_{min} \le v \le V_{max} \tag{6a}$$
$$T_{mmin} \le T_m \le T_{mmax} \tag{6b}$$
$$L_c \in [-1, 0, 1] \tag{6c}$$
$$d_{s,j} - \delta_{s,j} \ge 0 \tag{6d}$$
$$\begin{cases} v = 0 & if\ s_x = S_{tl} - S_{ts}\ and\ P = 0 \\ v \ne 0 & if\ s_x = S_{tl} - S_{ts}\ and\ P = 1 \end{cases} \tag{6e}$$
given:
$$x(T_s) = [V_s, 0],\ s_x(T_f) = S_{tl} \tag{6f}$$

where $T_s$ and $T_f$ are the initial and final time on the trip, respectively. $\zeta_e$ and $\zeta_t$ are coefficients that convert the electricity and time costs into their monetary counterpart, respectively. $V_s$ and $V_f$ are the vehicle's initial and final speeds, respectively. $T_{mmin} < 0$ and $T_{mmax} \ge 0$ are the maximum generation and propulsion torque of the motor, respectively. $d_{s,j}$ is the distance between the HV and the $j$th SV. $\delta_{s,j}$ is the minimum safe gap between the HV and the $j$th SV, which is defined by using an intelligent driver model (IDM) [32]. $S_{ts}$ is the standstill space between HV and the intersection stop line. $P$ is the traffic light indication, in the initial traffic light cycle $P = I_{in}$, following $P$ is determined from the traffic light model [16].

### C. Control Framework

The overtaking-enabled EAD optimal control problem is a complicated nonlinear time-varying optimization problem with multiscale objectives and various types of constraints, which is difficult to solve directly. Furthermore, the states of the SVs are dynamic changeable and difficult to forecast properly, which significantly affects the EAD control performance.

As a result, this paper proposes OEAC with a two-stage receding horizon control framework (see Fig. 3): the driving lane and speed are optimized in the first and second stages, respectively. Then, the overtaking-enabled EAD optimal control problem can be decomposed into driving lane planning and speed optimization problems, and solved by using MDP theory and PMP algorithm, respectively. Benefitting from the efficient optimization algorithm and receding horizon control framework, the OEAC is able to calculate the optimal vehicle driving trajectory in real time and adapt to dynamic traffic conditions.

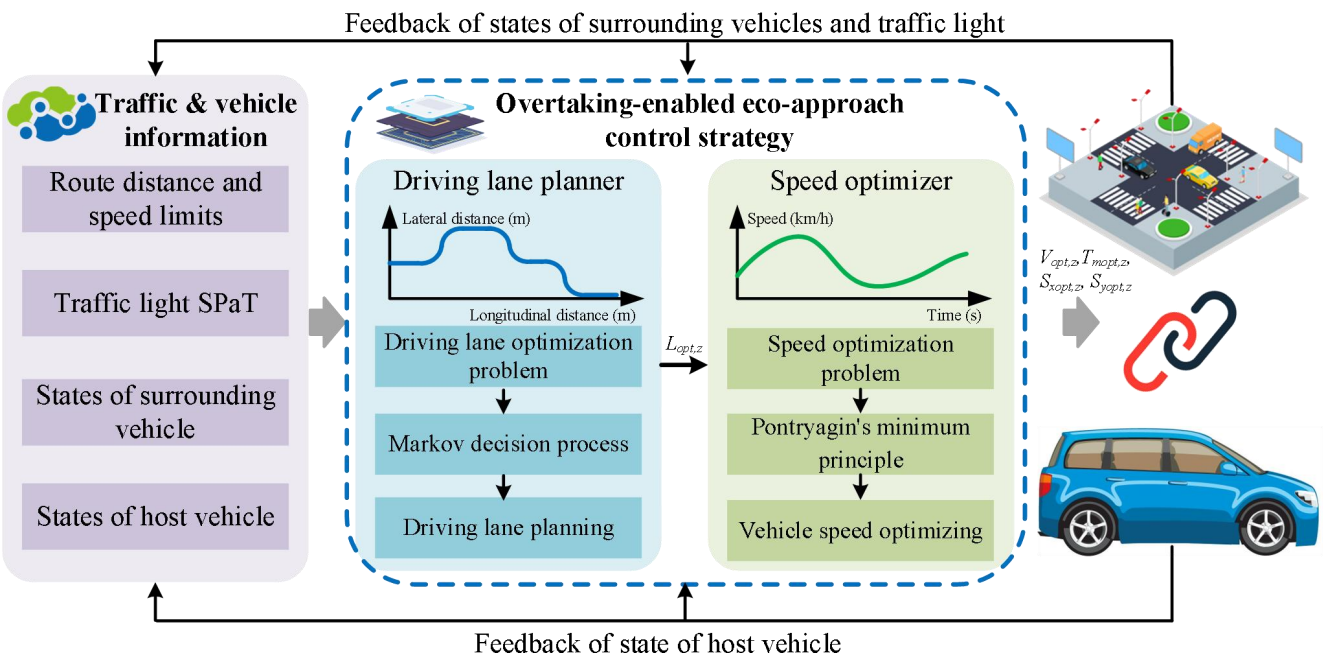

**Fig. 3.** The two-stage receding horizon control framework of the OEAC.

## IV. Overtaking-enabled Eco-approach Control Strategy

### A. Efficient Driving Lane Planning

*1) Driving Lane Optimization Problem:* Figure. 4 illustrates a tree graph to represent how the vehicle changes lanes while driving. The entire trip may be segmented into $N$ steps in the spatial domain, each of which necessitates choosing whether to change lanes or stay in the same lane. In lane-changing and lane-keeping operations, the travel period $\Delta t_z$, longitudinal distance interval $\Delta s_{x,z}$, and lateral distance interval $\Delta s_{y,z}$ are different. Note that the subscript $z = 1, 2, \ldots, N$ indicates the receding optimization step numbers.

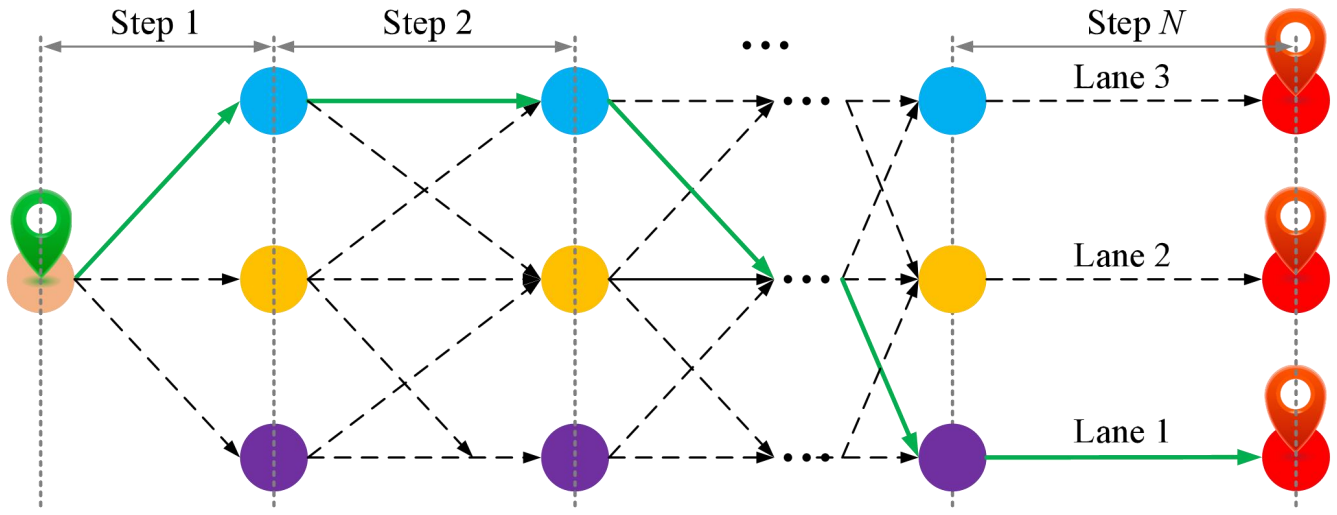

**Fig. 4.** A tree representation of the driving lane from God's eye point of view. Here the green solid and black dotted lanes donate the optimal and feasible driving lane sequences, respectively.

This study employs a receding horizon framework, assumes that the HV lane-changing operation continues once it begins until it enters the target lane. Given the state variable $x_{l,z} = [t_z, s_{x,z}, s_{y,z}]$ and control input $u_{l,z} = L_{c,z}$ . Then, the driving lane optimization problem (LOP) is formulated in (7),

$$\min_{u_{l,z} \in U} \mathcal{J}_l(x_{l,z}, u_{l,z}) = \sum_{z=1}^{N} (\zeta_e P_{b,z} + \zeta_t) \Delta t_z \tag{7}$$

subject to:
$$L_{c,z} \in [-1, 0, 1]$$
$$d_{s,j,z} - \delta_{s,j,z} \ge 0$$
$$x_{l,z+1} = \begin{bmatrix} t_{z+1} \\ s_{x,z+1} \\ s_{y,z+1} \end{bmatrix} = \begin{bmatrix} t_z + \Delta t_z \\ s_{x,z} + \Delta s_{x,z} \\ s_{y,z} + \Delta s_{y,z} \end{bmatrix}$$

where $\Delta t_z \in \{\Delta t_{c,z}, \Delta t_{k,z}\}$, $\Delta s_{x,z} \in \{\Delta s_{xc,z}, \Delta s_{xk,z}\}$, and $\Delta s_{y,z} \in \{s_{yc,z}, \Delta s_{yk,z}\}$ . $\Delta t_{c,z}$ and $\Delta t_{k,z}$ are time intervals of lane-changing and lane-keeping, respectively. $\Delta s_{xc,z}$ and $\Delta s_{xk,z}$ denote the longitudinal distances of lane-changing and lane-keeping, respectively. $s_{yc,z}$ and $\Delta s_{yk,z}$ are the lateral distances of lane-changing and lane-keeping, respectively. Then $\Delta t_{c,z} = \Delta s_{cx,z}/ V_{h,z}$ and $\Delta t_{k,z} = \Delta s_{kx,z}/ V_{h,z}$, where $V_{h,z}$ is the speed of HV at beginning of $z$th step. Note that $\Delta t_z$, $\Delta s_{xc,z}$, and $\Delta s_{yc,z}$ are related to lane-changing trajectories. Here we choose third-order polynomial based lane-changing trajectory model [33], which has the advantages of a closed form, continuous third derivative and smooth curvature.

*2) Safety Concerns during Lane-changing:* The operation of changing lanes offers a larger risk to safety than lane-keeping because the lateral displacement of the vehicle during lane-changing disrupts the flow of traffic in the adjacent lanes. The potential collision scenarios in the target lane can be categorized into two groups: rear-end collision caused by a slower preceding vehicle or moving faster rear vehicle; and side collision attributable to the side vehicle driving at the same speed as the HV. We define the flags that allow HV changes to the left or right lanes as $\psi_{cl}$ and $\psi_{cr}$, respectively. The safety lane-changing criteria are summarized in (8).

$$L_{c,z} = \begin{cases} -1 & \psi_{cl,z} = 1 \\ 0 & \psi_{cl,z} = 0 \text{ and } \psi_{cr,z} = 0 \\ 1 & \psi_{cr,z} = 1 \end{cases} \tag{8}$$

The lane-changing flags are determined by the safety clearance between HV and the nearest preceding, rear, or side vehicles, which is calculated by geometric methods [32], as given in (9),

$$\delta_{s,j,z} = H_s + T_h v_z + \frac{v_z(v_z - v_{s,j,z})}{2\sqrt{a_{min} a_{max}}} \tag{9}$$

where $H_s$ is the standstill distance between the HV and the nearest vehicle, $T_h$ is the safety time headway, $v_{s,j,z}$ is the speed of the nearest surrounding vehicles, and $a_{min}$ and $a_{max}$ are maximum deceleration and acceleration considering driving comfort, respectively.

*3) Efficient Driving Lane Planning using MDP Theory:* The movement of SV greatly influences the lane-changing decision of HV. Most research on lane-changing control assumes the movement of SV can be accurately accessed [20] or predicted using the SV movement model [24]. However, accurate prediction is more challenging because of the randomness behaviors of SV. In this context, this study employs MDP to find the optimal driving lane with consideration of uncertainties in SV's behaviors.

The LOP is a deterministic problem, which is transformed as an MDP [34], as indexed by $\mathcal{M} = \{\mathcal{W}, \mathcal{U}, \mathcal{P}, \mathcal{R}, \xi\}$, where $\mathcal{W}$ is the state space, $\mathcal{U}$ is the action space, $\mathcal{P}$ is the state transition probability from the current state to the next, $\mathcal{R}$ is the reward function, and $\xi$ is the discount factor. The following provides the definitions for all parameters.

The one-step reward $r(x_{l,z}, u_{l,z})$ is defined as driving cost, which is associated with a particular action, as listed in (10).

$$r(x_{l,z}, u_{l,z}) = -(\zeta_e P_{b,z} + \zeta_t)\Delta t_z \tag{10}$$

We define $\xi$ as an exponential decay factor so that long-term and short-term rewards can be balanced. Then, the $\mathcal{R}(x_{l,z}, u_{l,z})$ is calculated by (11).

$$\mathcal{R}(x_{l,z}, u_{l,z}) = r(x_{l,z+1}, u_{l,z+1}) + \xi r(x_{l,z+2}, u_{l,z+2}) + \xi^2 r(x_{l,z+3}, u_{l,z+3}) + \dots \tag{11}$$

The state transition probability is defined by a probability model $\mathcal{P}(x_{l,z+1} | x_{l,z}, u_{l,z}) \in \{\mathcal{P}_{cl,z}, \mathcal{P}_{cr,z}, \mathcal{P}_{lk,z}\}$ for the predicted optimal driving costs $C_{opt,z}$ and predicted actual driving costs (i.e., $C_{cl,z}$ and $C_{cr,z}$ for left and right turning, respectively), as listed in (12) – (14),

$$\mathcal{P}_{cl,z} = \begin{cases} \dfrac{C_{cl,z} - C_{opt,z}}{C_{cl,z} + C_{cr,z} + C_{lk,z} - 3C_{opt,z}} & \psi_{cl,z} = 1 \text{ and } C_{cl,z} < 2C_{opt,z} \\ 0 & \psi_{cl,z} = 0 \text{ or } C_{cl,z}(z) \geq 2C_{opt,z} \end{cases} \tag{12}$$

$$\mathcal{P}_{cr,z} = \begin{cases} \dfrac{C_{cr,z} - C_{opt,z}}{C_{cl,z} + C_{cr,z} + C_{lk,z} - 3C_{opt,z}} & \psi_{cr,z} = 1 \text{ and } C_{cr,z} < 2C_{opt,z} \\ 0 & \psi_{cr,z} = 0 \text{ or } C_{cr,z} \geq 2C_{opt,z} \end{cases} \tag{13}$$

$$\mathcal{P}_{lk,z} = 1 - \mathcal{P}_{cl,z} - \mathcal{P}_{cr,z} \tag{14}$$

where $\mathcal{P}_{cl,z}$, $\mathcal{P}_{cr,z}$, and $\mathcal{P}_{lk,z}$ are the probabilities of changing lane to the left, changing lane to the right, and lane-keeping, respectively. It should be noted that improvements in energy efficiency for eco-driving often do not surpass 100% [4], thus we set the probability of lane-changing to 0 when actual driving cost surpasses optimal driving cost by a factor of two.

The $C_{opt,z}$ is related to the feasible maximum time $T_{pmax,z}$ and minimum time $T_{pmin,z}$ to pass through the intersection. $T_{pmax,z}$ and $T_{pmin,z}$ can be calculated by considering the speed limits, vehicle position, and traffic light model [16]. Then, $C_{opt,z}$ is calculated by using (15),

$$C_{opt,z} = \min \begin{bmatrix} \sum_{t_s}^{T_{pmin,z}} (\zeta_e P_b + \zeta_t) T_{pmin,z} \\ \sum_{t_s}^{T_{pmin,z}+\Delta t} (\zeta_e P_b + \zeta_t)(T_{pmin,z} + \Delta t) \\ \vdots \\ \sum_{t_s}^{T_{pmax,z}} (\zeta_e P_b + \zeta_t) T_{pmax,z} \end{bmatrix} \tag{15}$$

where $t_s$ is the start time of the $z$th step, and $\Delta t = 0.1$s is the time interval. The actual driving costs depend on the driving operations and are affected by the movement of SVs. Taking into account the impacts of stochastic traffic environments, this study proposes a heuristic driving cost prediction method. The proposed method divides the challenging task of evaluating the driving cost within a long horizon into two parts: present driving cost and foresighted driving cost, as presented in (16) – (18),

$$C_{cl,z} = C_{clp,z} + C_{clf,z} \tag{16}$$

$$C_{cr,z} = C_{crp,z} + C_{crf,z} \tag{17}$$

$$C_{lk,z} = C_{lkp,z} + C_{lkf,z} \tag{18}$$

where $C_{clp,z}$, $C_{crp,z}$, and $C_{lkp,z}$ are the present driving costs for changing lane to the left, changing lane to the right, and lane-keeping, respectively, and $C_{clf,z}$, $C_{crf,z}$, and $C_{lkf,z}$ are the foresighted driving costs.

The present driving cost may be estimated more precisely since the traffic condition remains relatively constant for a short period. The cost of changing lanes to the left or right costs is the same, and the present driving costs can be calculated using (19) and (20).

$$C_{clp,z} = C_{crp,z} = \zeta_e P_b \Delta t_{c,z} + \zeta_t \Delta t_{c,z} \tag{19}$$

$$C_{lkp,z} = \zeta_e P_b \Delta t_{k,z} + \zeta_t \Delta t_{k,z} \tag{20}$$

In the calculation of foresighted driving costs, only the existing state of the preceding vehicle after the current driving lane selection operation is carried out, and the number of preceding vehicles is not evaluated. If the HV and the preceding vehicle do not collide, it is defined as lane-keeping mode; if the HV encounters the preceding vehicle upstream of the intersection, the timing and position of the encounter moment are estimated. In this situation, the HV will keep cruise-driving until encountering the preceding vehicle, and then follow the preceding vehicle. Therefore, the $C_{clf,z}$, $C_{crf,z}$, and $C_{lkf,z}$ are calculated using (21) – (23),

$$C_{clf,z} = \begin{cases} (\zeta_e P_b + \zeta_t)(t_{cdl,z} + t_{cfl,z}) & \vartheta = 1 \\ (\zeta_e P_b + \zeta_t)\dfrac{S_{tl} - \Delta s_{cx,z}}{V_{h,z}} & \vartheta = 0 \end{cases} \tag{21}$$

$$C_{crf,z} = \begin{cases} (\zeta_e P_b + \zeta_t)(t_{cdr,z} + t_{cfr,z}) & \vartheta = 1 \\ (\zeta_e P_b + \zeta_t)\dfrac{S_{tl} - \Delta s_{cx,z}}{V_{h,z}} & \vartheta = 0 \end{cases} \tag{22}$$

$$C_{lkf,z} = (\zeta_e P_b + \zeta_t)\frac{S_{tl} - \Delta s_{kx,z}}{V_{h,z}} \tag{23}$$

where $\vartheta$ is the existing state of the preceding vehicle, $\vartheta = 1$ denotes that existing the preceding vehicle on the target lane, and HV can encounter the preceding vehicle before it reaches the intersection stop line; on the contrary, $\vartheta = 0$ denotes the absence of the preceding vehicle or the preceding vehicle moving faster than HV. $t_{cdl,z}$ and $t_{cdr,z}$ are the cruise-driving time when HV catches up with the preceding vehicle by changing lanes to the left or right, respectively. $t_{cfl,z}$ and $t_{cfr,z}$ are the time of car-following after encounters preceding vehicles by changing lanes to the left or right, respectively.

The derivation for changing lanes to the left is demonstrated in the following; as the derivation for changing lanes to the right is identical, it is omitted. $t_{cdl,z}$ and $t_{cfl,z}$ are calculated using (24) and (25), respectively,

$$t_{cdl,z} = \begin{cases} \dfrac{d_{hp,z}}{V_{h,z} - V_{p,z}} & \text{if } T_{cpl,z} > \dfrac{d_{hp,z}}{V_{h,z} - V_{p,z}} \\ 0 & \text{others} \end{cases} \tag{24}$$

$$t_{cfl,z} = \begin{cases} \dfrac{S_{tl} - \Delta s_{xc,z} - V_{h,z} t_{cdl,z}}{V_{p,z}} & \text{if } T_{cpl,z} > \dfrac{d_{hp,z}}{V_{h,z} - V_{p,z}} \\ 0 & \text{others} \end{cases} \tag{25}$$

where $d_{hp,z}$ is the distance between HV and the preceding vehicle, $T_{cpl,z} = (S_{tl} - s_{xp,z} - V_{p,z}\Delta t_{c,z})/V_{p,z}$ is the time for the preceding vehicle to pass through the intersection. $V_{p,z}$ and $s_{xp,z}$ are the speed and initial position of the preceding vehicle, respectively.

Then, the optimal solution of value function $V(x_{l,z})$ and control policy $\pi(x_{l,z})$ satisfies (26) and (27).

$$V(x_{l,z}) = \max_{u_{l,z}\in U} \mathcal{R}(x_{l,z}, u_{l,z}) + \sum_{x_{l,z+1}} \mathcal{P}(x_{l,z+1} | x_{l,z}, u_{l,z}) \mathcal{V}(x_{l,z+1}) \tag{26}$$

$$\pi(x_{l,z}) = \operatorname*{argmax}_{u_{l,z}\in U} \mathcal{R}(x_{l,z}, u_{l,z}) + \sum_{x_{l,z+1}} \mathcal{P}(x_{l,z+1} | x_{l,z}, u_{l,z}) \mathcal{V}(x_{l,z+1}) \tag{27}$$

Dynamic programming is utilized to resolve the MDP problem [35] and determine the optimal driving lane $\mathcal{L}_{opt,z}$.

*B. Energy-saving Speed Optimization*

*1) Speed Optimization Problem:* To get a real time analytical solution, the speed optimization problem (SOP) is formulated, assuming that the driveline and battery system efficiencies are constant, i.e., $\eta_t = 0.9$ and $\eta_b = 0.9$ [36]. Since a vehicle travels at modest speeds in urban traffic, air resistance is ignored [37]. In addition, as urban roads are often level, this paper does not evaluate the impact of road slopes. To ensure a viable initial state, the state constraints are inactive at the beginning, then the SOP can be formulated, as provided in (28),

$$\min_{u_s\in U} \mathcal{J}_s(x, u_s) = \int_{t_s}^{t_f} \left\{ (\varpi v u_s + \kappa u_s^2)\eta_b + \frac{P_a}{\eta_b} \right\} \zeta_e dt \tag{28}$$

subject to: $\dot{s_x} = v$, $\dot{v} = \frac{u_s i_g i_0 \eta_t}{m\delta r_w} - \frac{gf}{\delta}$, (6a), (6b)

given: $x(t_s) = [v_s, s_s]^T$, $s(t_f) = s_f$

where $u_s$ is the control input of SOP, that is, $u_s = T_m$. $t_s$ is the initial time of the current step, and $t_f$ is the estimated intersection passing time calculated using (24) and (25).

*2) Analytical Solution using PMP Algorithm:* Different types of constraints are included in the SOP. According to the PMP solution criteria, the unconstrained solution is formed first, which is the basis to derive a constrained solution. The *Hamiltonian* in the unconstrained condition is given as (29),

$$H(x, u_s, t) = \left\{ (\varpi v u_s + \kappa u_s^2)\eta_b + \frac{P_a}{\eta_b} \right\} \zeta_e + \phi_p v + \frac{\phi_s}{\delta}\left( \frac{u_s i_g i_0 \eta_t}{m r_w} - gf \right) \tag{29}$$

where $\phi_p$ and $\phi_s$ are the position and speed co-state variables, respectively. Following the *Euler-Lagrange* theory, the necessary conditions for optimality are given in (30) – (32).

$$\frac{\partial H}{\partial u_s} = (\varpi v + 2\kappa u_s)\eta_b \zeta_e + \frac{\phi_s i_g i_0 \eta_t}{m\delta r_w} = 0 \tag{30}$$

$$\frac{\partial H}{\partial x} = -\dot{\phi}_p = 0 \tag{31}$$

$$\frac{\partial H}{\partial v} = -\dot{\phi}_s = \varpi \eta_b \zeta_e u_s + \phi_p = 0 \tag{32}$$

Since $\dot{\phi}_p = 0$ yields $\phi_p$, $v = v_s + \left(\frac{u_s i_g i_0 \eta_t}{m\delta r_w} - \frac{gf}{\delta}\right) t$, and $u_s = -\frac{\phi_p}{\varpi \eta_b \zeta_e}$ (from (32), $\dot{\phi}_s = 0$ when $t = t_f$), the optimal control input $u_s^*$, speed $v^*$, and longitudinal position $s_x^*$ can be calculated by using vehicle dynamics and (30) – (32), as shown in (33) – (35),

$$u_s^* = \Lambda_a t + \Lambda_b \tag{33}$$

$$v^* = \frac{\Lambda_a i_g i_0 \eta_t}{2m\delta r_w} t^2 + \left(\frac{\Lambda_b i_g i_0 \eta_t}{m r_w} - gf\right)\frac{t}{\delta} + e_v \tag{34}$$

$$s_x^* = \frac{\Lambda_a i_g i_0 \eta_t}{6m\delta r_w} t^3 + \left(\frac{\Lambda_b i_g i_0 \eta_t}{m r_w} - gf\right)\frac{t^2}{2\delta} + e_v t + e_s \tag{35}$$

with

$$\Lambda_a = \frac{1}{2\kappa\delta}\left(\varpi g f + \frac{\phi_p i_g i_0 \eta_t}{m r_w \eta_b \zeta_e}\right)$$

$$\Lambda_b = -\frac{1}{2\kappa}\left(\varpi v_s + \frac{\phi_s i_g i_0 \eta_t}{m\delta r_w \eta_b \zeta_e}\right)$$

where $e_v$ and $e_s$ are the constants of integration, which are determined by boundary conditions.

The optimal solution, which respects the speed and control constraints, can be obtained from the unconstrained optimal solution. Since the speed limits are pure state inequality constraints that are independent of the control input, we adjoin them in the *Hamiltonian* using the indirect method [38], which adjoins the $p^{\text{th}}$ order state constraint indirectly up until the control input is explicitly present. As a result, the explicit control input for the speed limits has to be included in one differential computation. As such, we define $A^{(p)}(x, u_s, t)$ as the adjoining function of speed and control input constraints, as expressed in (36) – (39).

$$A_1^{(1)}(x, u_s, t) = \frac{1}{\delta}\left(\frac{u_s i_g i_0 \eta_t}{m r_w} - gf\right) \tag{36}$$

$$A_2^{(1)}(x, u_s, t) = -\frac{1}{\delta}\left(\frac{u_s i_g i_0 \eta_t}{m r_w} - gf\right) \tag{37}$$

$$A_3^{(0)}(x, u_s, t) = u_s - T_{mmax} \tag{38}$$

$$A_4^{(0)}(x, u_s, t) = T_{mmin} - u_s \tag{39}$$

The resulting *Lagrangian* is formed as (40),

$$\begin{aligned} L(x, u_s, t) = {} & \\ & H(x, u_s, t) + \varphi_1 A_1^{(1)}(x, u_s, t) + \varphi_2 A_2^{(1)}(x, u_s, t) \\ & + \xi_1 A_3^{(0)}(x, u_s, t) + \xi_2 A_4^{(0)}(x, u_s, t) \end{aligned} \tag{40}$$

where $\varphi$ and $\xi$ are the co-states of speed and control input constraints. Note that the values of $\varphi$ and $\xi$ are always zero when the speed and control constraints are not active, but they are not zero when the constraints are triggered.

Considering $A^{(p)}(x, u_s, t) \leq 0$ does not prevent the speed and torque trajectories from exceeding the boundaries. Then, the tangency conditions [39] of $p^{\text{th}}$ order state constraints at the junction time must be added, as defined in (41),

$$B(x, u_s, t) = \begin{bmatrix} v(\tau) - V_{max} \\ V_{min} - v(\tau) \\ u_s(\tau) - T_{mmax} \\ T_{mmin} - u_s(\tau) \end{bmatrix} = 0 \tag{41}$$

where $\tau$ is the entry time of the constraint.

The jump conditions derived from interior-point constraints at each relevant moment are defined in (42) – (44),

$$\phi_p^*(\tau^-) - \phi_p^*(\tau^+) = \sum_{j=0}^{p-1} \pi_j G_{x^*}^{(j)}(x^*, \tau) \tag{42}$$

$$\phi_s^*(\tau^-) - \phi_s^*(\tau^+) = \sum_{j=0}^{p-1} \pi_j G_{x^*}^{(j)}(x^*, \tau) \tag{43}$$

$$H(\tau^-) - H(\tau^+) = \sum_{j=0}^{p-1} \pi_j G_t^{(j)}(x^*, \tau) \tag{44}$$

where $\pi_j$ is a *Lagrange* multiplier determined to satisfy constraints, and $\tau^-$ and $\tau^+$ are the time instance just before and after time $\tau$. Our previous research has revealed that the position co-state, speed co-state, *Hamiltonian*, and control input are continuous at $t = \tau$ when the states and control input constraints are triggered independently [40].

Let $\mathcal{F}$ be the constraint violated flag set, consisting of four bits representing the states of maximum speed, minimum speed, maximum control input, and minimum control constraints, respectively. A value of 1 indicates that the constraint has been activated, while a value of 0 indicates that it has not. For example, $\mathcal{F} = [1, 0, 0, 0]$ indicates that only the maximum speed constraint is activated. Then, the closed-form analytical solution of each state and control input constraints activation at $t \in [t_1, t_2]$ for the optimal control input $u_s^*$, states ($v^*$ and $s_x^*$), and co-state variables ($\varphi_1$, $\varphi_2$, $\xi_1$, $\xi_2$) are calculated by (45) – (48).

$$u_s^* = \begin{cases} \dfrac{mgf\delta r_w}{i_g i_0 \eta_t} & \mathcal{F} = [1, 0, 0, 0] \\ \dfrac{mgf\delta r_w}{i_g i_0 \eta_t} & \mathcal{F} = [0, 1, 0, 0] \\ T_{mmax} & \mathcal{F} = [0, 0, 1, 0] \\ T_{mmin} & \mathcal{F} = [0, 0, 0, 1] \end{cases} \tag{45}$$

$$\begin{cases} \varphi_1 = -\dfrac{m\delta r_w \eta_b \zeta_e(\varpi v + 2\kappa u_s)}{i_g i_0 \eta_t} - \phi_s & \mathcal{F} = [1, 0, 0, 0] \\ \varphi_2 = \dfrac{m\delta r_w \eta_b \zeta_e(\varpi v + 2\kappa u_s)}{i_g i_0 \eta_t} + \phi_s & \mathcal{F} = [0,1, 0, 0] \\ \xi_1 = -\eta_b \zeta_e(\varpi v + 2\kappa u_s) - \dfrac{\phi_s i_g i_0 \eta_t}{m\delta r_w} & \mathcal{F} = [0, 0,1, 0] \\ \xi_2 = \eta_b \zeta_e(\varpi v + 2\kappa u_s) + \dfrac{\phi_s i_g i_0 \eta_t}{m\delta r_w} & \mathcal{F} = [0, 0, 0, 0] \end{cases} \tag{46}$$

$$v^* = \begin{cases} V_{max} & \mathcal{F} = [1, 0, 0, 0] \\ V_{min} & \mathcal{F} = [0, 1, 0, 0] \\ \left(\dfrac{i_g i_0 \eta_t T_{mmax}}{m r_w} - gf\right)\dfrac{t}{\delta} + e_v & \mathcal{F} = [0, 0, 1, 0] \\ \left(\dfrac{i_g i_0 \eta_t T_{mmin}}{m r_w} - gf\right)\dfrac{t}{\delta} + e_v & \mathcal{F} = [0, 0, 0, 1] \end{cases} \tag{47}$$

$$s_x^* = \begin{cases} V_{max}(t_2 - t_1) + e_s & \mathcal{F} = [1, 0, 0, 0] \\ V_{min}(t_2 - t_1) + e_s & \mathcal{F} = [0, 1, 0, 0] \\ \left(\dfrac{i_g i_0 \eta_t T_{mmax}}{m r_w} - gf\right)\dfrac{t^2}{2\delta} + e_s & \mathcal{F} = [0, 0, 1, 0] \\ \left(\dfrac{i_g i_0 \eta_t T_{mmin}}{m r_w} - gf\right)\dfrac{t^2}{2\delta} + e_s & \mathcal{F} = [0, 0, 0, 1] \end{cases} \tag{48}$$

Typically, the unconstrained optimal solution is calculated first, then resolving the SOP by combining the solutions to the constrained and unconstrained problems if the unconstrained optimal solution violates any of the constraints. The procedure is continued until no more constraints are triggered by the solution. However, the large number of repeated calculations raises the computational cost. In this context, the active conditions of state and control input constraints are implemented based on the findings of Dong *et.al* [40] and Mahbub *et.al* [41], to eliminate some constraints. For example, the control constraints only exist in the first phase of the optimal control sequence. Therefore, only state constraints rather than control input constraints may be activated later if the initial phase focuses on the unconstrained optimal solution.

The optimal control sequence depends on the order in which the vehicle state and control constraints are activated. For each case, the optimal control input, states, and entry time yield a set of two-point boundary value problem algebraic equations, which must be simultaneously solved while taking all constraints into consideration. Finally, a series of control input and states variable (i.e., $u_s^*$, $v^*$, and $s_x^*$) within the prediction horizon can be obtained by using PMP in each receding optimization.

In summary, in the $z$th receding optimization, the optimal control $\mathcal{T}_{opt,z}$, speed $\mathcal{V}_{opt,z}$, longitudinal position $\mathcal{S}_{xopt,z}$, lateral position $\mathcal{S}_{yopt,z}$ are derived based on the $\mathcal{L}_{opt,z}$. During the lane-changing process, $\mathcal{V}_{opt,z} = V_{h,z}$, $\mathcal{T}_{opt,z} = mgfr_w/i_g i_0 \eta_t$, and $\mathcal{S}_{xopt,z}$ and $\mathcal{S}_{yopt,z}$ are calculated by using vehicle kinematics and lane-changing trajectory models. In the lane-keeping process, the lateral position is not changed, therefore $\mathcal{S}_{yopt,z} = 0$. The derived optimal control input and state variables at the first step are then applied, that is, $\mathcal{T}_{mopt,z} = u_s^*(1)$, $\mathcal{V}_{opt,z} = v^*(1)$, and $\mathcal{S}_{xopt,z} = s_x^*(1)$.

## V. Simulation and Results

### A. Simulation Setup

The vehicle parameters are taken from the Chery Little-Ant, as listed in Table I. The coefficients that convert the energy consumption and travel time into the monetary counterpart are 0.12 USD/kWh and 24 USD/hour, respectively, according to the study in [42]. The traffic system is simulated using the real-world urban route as shown in Fig. 5, i.e., the intersection of Shuanglong Avenue and Jiyin Avenue, Nanjing, China, with the parameters listed in Table II. The IDM model is employed as the car-following model to imitate the driving behavior of vehicles, and the LC2013 model is used for describing lane-changing decisions [43]. The simulation is performed in MATLAB (version 9.13, 2022b) on a PC with Intel Core i9-12900k @ 3.20GHz CPU and 64GB RAM, the sampling time is 0.01s in simulation.

### B. Benchmark Strategy

To evaluate the performance of the OEAC, the constant speed (CS) and READ strategies are used as benchmarks. In a free-driving situation, the CS strategy generally maintains a constant speed for the vehicle [12]. In addition, READ generates an energy-efficient speed profile using the PMP algorithm while only considering lane-keeping operation [40]. When a preceding vehicle appears in front of the HV, the car-following strategy is activated in both CS and READ to keep a safe distance. The car-following driving behavior is modeled by the IDM.

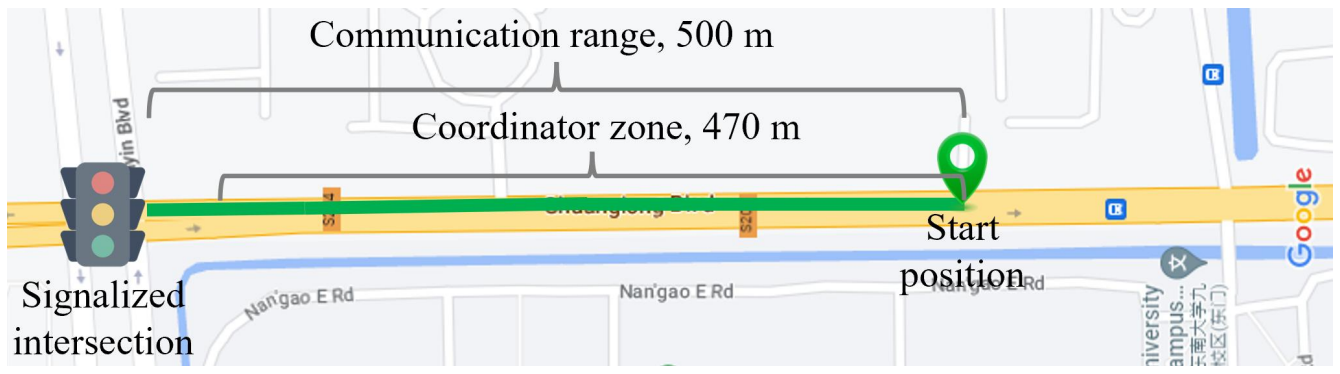

**Fig. 5.** Map of test route.

TABLE I
VEHICLE PARAMETERS.

| Parameter | Value | Parameter | Value |
|---|---|---|---|
| Air density $\rho$ | 1.206kg/m$^3$ | Tire radius $r_w$ | 0.280m |
| Air drag coefficient $C_D$ | 0.3 | Transmission ratio $i_g$ | 2.80 |
| Front area $A$ | 2.02m$^2$ | Motor loss coefficient $M_2$ | 0.873 |
| Final drive ratio $i_0$ | 3.789 | Vehicle mass $m$ | 1005kg |
| Rolling resistance $f$ | 0.015 | Auxiliary power $P_a$ | 300W |
| Rotational inertia $\delta$ | 1.022 | Gravity acceleration $g$ | 9.8m/s$^2$ |
| Maximum acceleration $a_{max}$ | 2m/s$^2$ | Motor maximum propulsion torque $T_{mmax}$ | 69Nm |
| Maximum deceleration $a_{min}$ | -2m/s$^2$ | Motor maximum generation torque $T_{mmin}$ | -61Nm |

TABLE II
ROADWAY NETWORK AND TRAFFIC PARAMETERS.

| Parameter | Value | Parameter | Value |
|---|---|---|---|
| Location of stop line $S_{tl}$ | 500m | Width of lane $D_w$ | 3.5m |
| Communication range $D_{cr}$ | 500m | Time of red signal $T_{gr}$ | 51s |
| Coordinator zone $D_{lr}$ | 470m | Time of green signal $T_{gr}$ | 35s |
| Maximum speed $V_{max}$ | 70km/h | Standstill distance $H_s$ | 2m |
| Minimum speed $V_{min}$ | 20km/h | Safety time headway $T_h$ | 1.25s |
| Number of lanes $N_l$ | 3 | Road slope $\theta$ | 0 |

### C. Simulation Results

*1) Performance Verification in Stochastic Traffic Scenarios:* To evaluate the driving cost reduction effectiveness of OEAC, a stochastic simulation is conducted. The HV is initially located in the middle lane, and 10000 individual simulation trials are conducted with constant $S_{tl}$, $D_{cr}$, $D_{lr}$, $V_{min}$, $V_{max}$, $T_{gr}$, and $T_{re}$. The parameters of the HV initial speed, traffic light initial states (i.e., $I_{in}$ (0 or 1) and $T_{ls}$ (1–$T_{gr}$ or 1–$T_{re}$)) traffic flow conditions (vehicle density (0 – 240 veh/km), and initial position and speed of SV) are all randomized in the simulation for a thorough investigation. Note that $V_s$ is equal to the average traffic flow speed and is related to vehicle density.

The heat maps of the driving cost reduction of stochastic tests are depicted in Figs. 6 and 7. We point out that each grid of the heat map represents the average improvement of the different traffic scenarios with various speeds and positions of SVs. Table III compares the driving cost reduction and computational time of the OEAC in each step.

Figs. 6 and 7 display the distribution of driving cost reduction for the OEAC under various traffic conditions. The minimum and maximum driving cost reduction of the OEAC are 0% and 59.97%, respectively, when compared to the CS. In comparison to the READ, the OEAC can cut the driving cost usage by up to 59.02%. On average, the OEAC performs significantly better than the CS and READ, achieving a driving cost reduction of 20.91% and 5.62%, respectively. The findings indicate that the OEAC is remarkably effective in urban traffic. Although the driving cost reduction is similar to that of CS and READ in some scenarios, in most situations, the OEAC can effectively lower the driving cost. In addition, the initial states of traffic lights and traffic flow conditions have an impact on how well the OEAC performs in terms of reducing driving costs. Further investigation into this will be conducted below.

Furthermore, as shown in Table III, the average and maximum computational time in each step of the OEAC are 0.2ms and 9.3ms, respectively, both of which are less than the sampling time (10ms). These results show that the proposed OEAC is computationally efficient and has the potential to be deployed in real time.

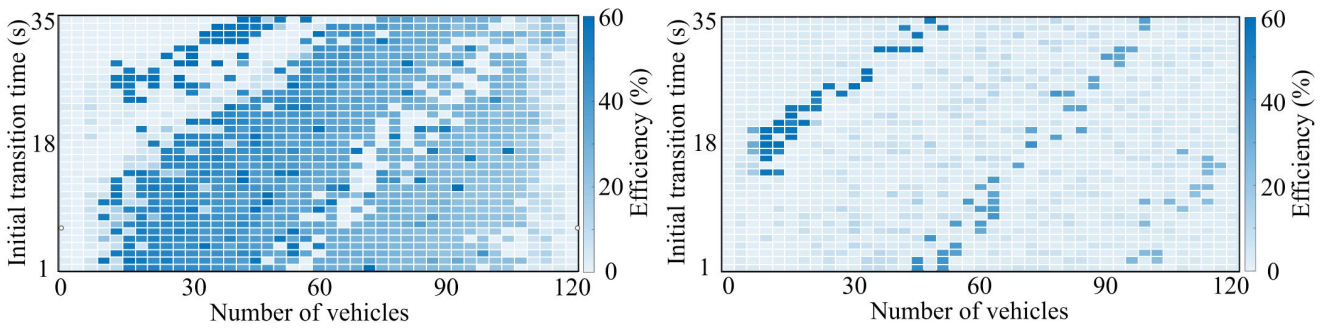

(a) OEAC versus CS. (b) OEAC versus READ.

**Fig. 6.** Heat map of driving cost reduction of stochastic simulations, here the initial indication of the traffic light is green.

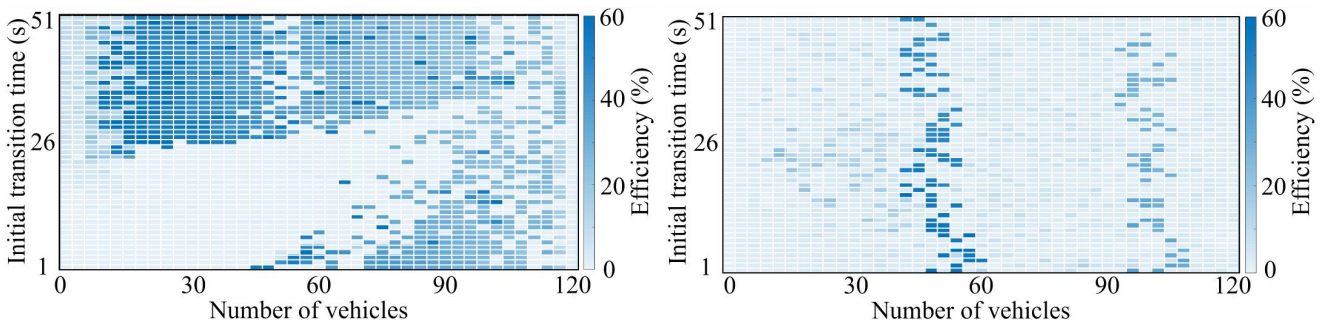

(a) OEAC versus CS. (b) OEAC versus READ.

**Fig. 7.** Heatmap of driving cost reduction of stochastic simulations, here the initial indication of the traffic light is red.

TABLE III
RESULTS OF STOCHASTIC SIMULATION.

| Strategies | Driving cost reduction | | | Computational time in each step of OEAC |
|---|---|---|---|---|
| | Minimum | Maximum | Average | |
| OEAC versus CS | 0% | 59.97% | 20.91% | 0.2ms (average) 9.3ms (maximum) |
| OEAC versus READ | 0% | 59.02% | 5.62% | |

*2) Typical Cases Analysis:* The stochastic simulation results allow for the selection of three typical cases with different vehicle density conditions: Cases A (free-flow), B (moderate-flow), and C (congested-flow). Table IV lists the parameters of three cases. A typical case analysis was then conducted to show the vehicle driving trajectories of three control strategies, enabling the further demonstration of the driving cost reduction by the OEAC. Figs. 8 – 10 show the vehicle speed, position, driving lane, and acceleration trajectories in three cases. Table V provides the vehicle driving cost, travel time, and energy consumption data for the OEAC and benchmark strategies. For a fair comparison, the energy consumption includes kinetic energy change between the starting and final positions and battery energy usage.

As shown in Fig. 8 and Table V, all three strategies enable HV to reach the destination without the need for lane change operations (Fig. 8(c)), while also remaining unaffected by preceding vehicles. As a consequence, both OEAC and READ strategies yield the same solution. Although OEAC and READ have longer travel times compared to CS, they save vehicle energy consumption. When compared to the CS strategy, the OEAC and READ can reduce the driving cost by 4.17%. This is due to the capacity of OEAC and READ to leverage traffic light information to optimize smoothly changing speed and acceleration (see Figs. 8(a) and 8(d)) so that the HV passes through the signalized intersection at the least driving cost. According to the findings, OEAC offers no benefit over READ and provides only slight driving cost reduction over CS in the free-flow traffic condition due to the nature of this situation.

Fig. 9 illustrates that both the CS and READ need the controlled HV to pass through the signalized intersection at the second traffic light green signal window (see Fig. 9(b)) because the preceding vehicles hinder the movement of HV. As a result, see Table V, the driving cost of CS and READ increases as vehicle energy usage and travel time rise. In contrast, the OEAC changes the driving lane from lane 2 to lane 3 (see Fig. 9(d)), allowing the HV to pass through the signalized intersection during the first traffic light green signal window (see Fig. 9(c)), thereby reducing the driving cost. Compared to the CS and READ, the OEAC effectively optimizes the vehicle driving lane and speed in moderate-flow conditions, resulting in a decrease of 55.93%/55.89%, 36.43%/31.46%, and 54.12%/53.01% in travel time, energy consumption, and driving cost, respectively. This reveals that the energy and travel time of the HV is more affected by the preceding vehicles in moderate-flow. The OEAC performs effectively in moderate-flow conditions by optimizing the vehicle driving lane and speed, lowering vehicle energy consumption and travel time to reduce driving costs.

TABLE IV
PARAMETERS OF THREE TYPICAL CASES.

| Parameters | Case A | Case B | Case C |
|---|---|---|---|
| Vehicle density | 8veh/km | 108veh/km | 240veh/km |
| Average speed of traffic flow | 70km/h | 48km/h | 20km/h |
| Initial transition time of traffic light | 35s | 20s | 20s |
| Initial indication of traffic light | Green | Red | Green |

TABLE V
SIMULATION RESULTS OF THREE TYPICAL CASES.

| Parameters | | Case A | Case B | Case C |
|---|---|---|---|---|
| Driving cost | CS | 0.24$ | 0.85$ | 1.37$ |
| | READ | 0.23$ | 0.83$ | 1.36$ |
| | OEAC | 0.23$ | 0.39$ | 1.36$ |
| Travel time | CS | 25.80s | 114.30s | 191.71s |
| | READ | 30.02s | 114.20s | 192.01s |
| | OEAC | 30.02s | 50.37s | 192.01s |
| Energy consumption | CS | 202.95kJ | 305.41kJ | 279.17kJ |
| | READ | 162.59kJ | 283.26kJ | 263.73kJ |
| | OEAC | 162.59kJ | 194.15kJ | 263.73kJ |

In congested-flow conditions, as shown in Fig. 10, the HV with OEAC passes through the signalized intersection without lane-changing operation. Because there are more vehicles on the road, the average speed of the traffic flow is slower, and the spacing between vehicles is smaller; hence, there is no appropriate lane-changing opportunity for HV to reduce travel time and energy consumption. The strategies OEAC and READ have identical driving trajectories, so their energy consumption, travel time, and driving cost are identical (see Table V).

*3) Sensitive Analysis for the OEAC:* According to the findings of the stochastic traffic scenarios simulation and typical case analysis, the effectiveness of OEAC is influenced by many factors. This subsection provides a comprehensive investigation of various factors that impact the performance improvement of OEAC, particularly concerning traffic flow conditions and initial states of traffic lights.

In free-flow conditions, HV is unaffected by the preceding vehicles in all lanes, then the probability of lane-changing of HV is 0. The OEAC and READ have the same performance and are better than CS. As shown in Fig. 9(b), the degree of OEAC's superiority over CS depends on the traffic light's initial states. For example, if the average traffic flow speed is 70km/h, HV arrives at the intersection at the 25.71s under CS control. As a result, if the initial indication is red, the HV has a stop-and-go operation at the intersection under CS control if the initial transition time exceeds 25.71s. However, OEAC can avoid the stop-and-go operation by optimizing the vehicle speed. Because the operation of stop-and-go consumes more energy, the energy efficiency of OEAC is preferable to CS. In addition, the travel time difference between OEAC and CS is small, as both are impacted by the initial traffic light states. Therefore, the advantage of OEAC over CS in driving cost reduction is mostly from the reduced energy usage brought about by the optimized free-driving speed. Consequently, in the free-flow condition, the effectiveness of OEAC results is due to speed optimization and is sensitive to traffic light initial states (see Fig. 6(a) and 7(a)).

In congested-flow conditions, the effectiveness of OEAC is similar to CS and READ as there is limited safety clearance for lane-changing due to slower preceding vehicles. In contrast to the free-flow condition, the effectiveness of OEAC over CS in the congested-flow condition is sensitive to the SV condition, because CS, READ, and OEAC strategies are all run in car-following mode (see Fig. 10(b)). The advantage of OEAC over CS in driving cost reduction is primarily due to the optimized car-following speed, which leads to lower energy consumption. Thus, the driving cost reduction of OEAC over CS is more spread with different traffic light initial states, as shown in Fig. 6(a) and 7(a).

In moderate-flow conditions, the HV has more possibilities to change lanes, hence the advantage of OEAC is obvious in reducing driving costs. Under the control of OEAV, HV can choose the lane that allows for the fastest passage through the signalized intersection while optimizing the driving speed to reduce travel time and energy consumption. As shown in Figs. 6 and 7, OEAC exhibits a greater improvement in driving cost reduction compared to CS and READ in the 20 – 220veh/km range of vehicle density. In addition, the improvements are more uniformly distributed, as the probability of lane-changing is sensitive to both the SV condition and the initial states of traffic lights.

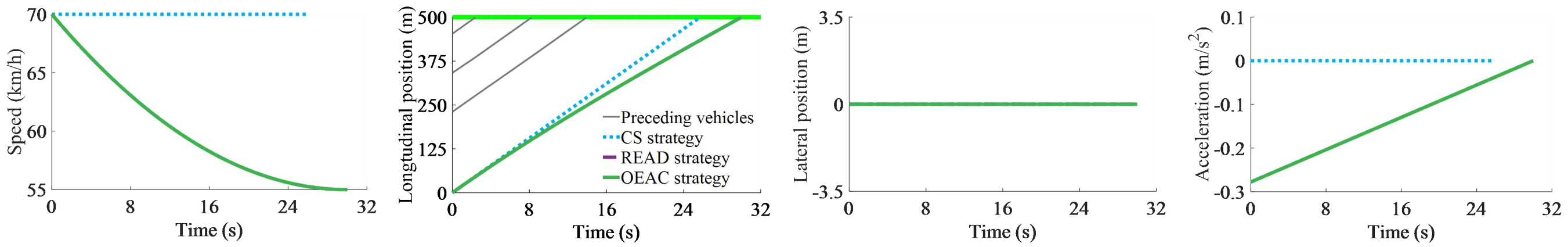

(a) Speed. (b) Longitudinal position (Lane 2). (c) Lateral position. (d) Acceleration.

**Fig. 8.** Simulation results of Case A. Here, the HV keeps driving in lane 2 of CS, READ, and OEAC strategies.

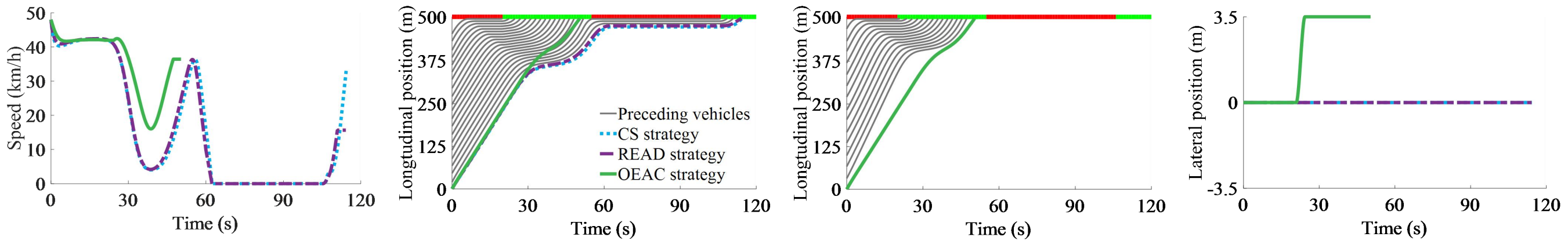

(a) Speed. (b) Longitudinal position (Lane 2). (c) Longitudinal position (Lane 3). (d) Lateral position.

**Fig. 9.** Simulation results of Case B. Here, the HV keeps driving in lane 2 firstly and changes to lane 3 at 20.91s of OEAC strategy, and keeps driving in lane 2 of CS and READ strategies.

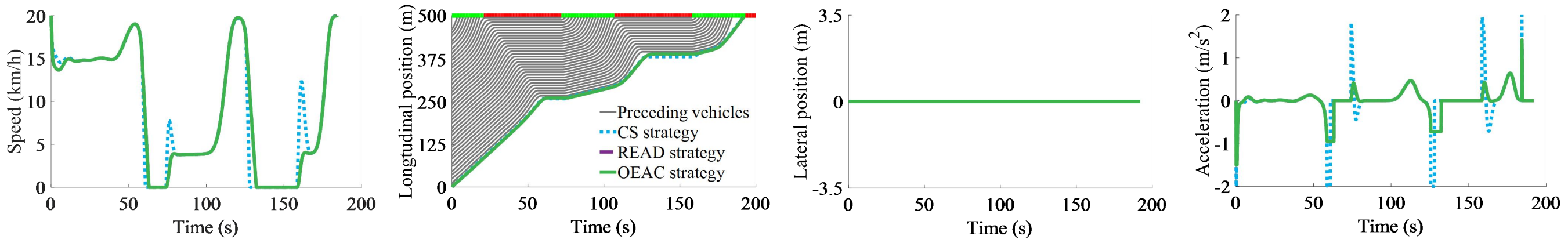

(a) Speed. (b) Longitudinal position (Lane 2). (c) Lateral position. (d) Acceleration.

**Fig. 10.** Simulation results of Case C. Here, the HV keeps driving in lane 2 of CS, READ, and OEAC strategies.

## Conclusion

This study proposes an overtaking-enabled EAD control strategy for CAVs that combines driving lane planning and speed optimization to efficiently reduce driving costs. An overtaking-enabled EAD optimal control problem is formulated with unified monetary counterpart objectives of energy consumption and travel time while accounting for the traffic light and preceding vehicle constraints. The OEAC utilizes the receding horizon two-stage control framework to solve overtaking-enabled EAD optimal control problems in real time, resulting in an optimal driving trajectory. In the first stage, the MDP theory is adopted to plan an efficient driving lane while accounting for the uncertain disturbance of preceding vehicles. In the second stage, the speed trajectory is optimized with the PMP algorithm to minimize driving costs with the minimized computational cost.

The stochastic traffic scenarios simulation and typical case study are investigated to verify the effectiveness of the OEAC. The results demonstrate that the OEAC outperforms CS and READ strategies in various traffic environments, with an average improvement of 20.91% and 5.62%, respectively. Furthermore, the sensitivity analysis findings manifest that OEAC's capability of lowering driving costs is mainly affected by the initial states of traffic lights in free flow, SV conditions in moderate flow, and both SV conditions and the initial states of traffic lights in congested flow.